\documentclass[10pt]{article}
\usepackage[utf8x]{inputenc}
\usepackage{graphicx}
\usepackage{amsmath,amssymb}
\usepackage{fullpage}
\usepackage{wrapfig}

\newtheorem{defin}{Definition}
  
\newtheorem{theo}[defin]{Theorem}
  \newenvironment{theorem}{\begin{theo} \sl}{\end{theo}}
\newtheorem{lem}[defin]{Lemma}
  \newenvironment{lemma}{\begin{lem} \sl}{\end{lem}}
\newtheorem{coro}[defin]{Corollary}
  
\newenvironment{myproof}{\emph{Proof.}}{\hfill $\Box$ \medskip\\}

\newcommand{\etal}{{\emph{et~al.}}}
\newcommand{\from}[1]{{\emph{\textbf{(#1)}}}}

\newenvironment{shortitemize}
    {\begin{itemize}\setlength{\itemsep}{-5pt}}
    {\end{itemize}}

\title{A History of Flips in Combinatorial Triangulations\thanks{As flips are a topic close to Ferran Hurtado's heart, we would like to dedicate this article to him on the occasion of his 60th birthday.}\addtocounter{footnote}{+2}}

\author{Prosenjit Bose\thanks{School of Computer Science, Carleton University. Research supported in part by NSERC. Email: \texttt{jit@scs.carleton.ca, sander@cg.scs.carleton.ca}.}
\and
\addtocounter{footnote}{-1}
Sander Verdonschot\footnotemark
}
\date{}

\begin{document}

\maketitle
\begin{abstract}
Given two combinatorial triangulations, how many edge flips are necessary and sufficient to convert one into the other? This question has occupied researchers for over 75 years. We provide a comprehensive survey, including full proofs, of the various attempts to answer it.
\end{abstract}

\section{Introduction}
\label{sec:introduction}

A triangulation is a simple planar graph where every face including the outer face is a cycle of length 3.  In any triangulation, an edge $e=xy$ is adjacent to two faces: $xya$ and $xyb$. An edge flip consists of deleting the edge $e$ from the triangulation and adding the unique edge $e'=ab$ to the graph such that it remains a triangulation. In other words, the edge $e$ is flippable provided that $ab$ is not currently an edge of the triangulation. If the vertices have fixed coordinates in the plane, the restriction that the new edge may not introduce any crossings is usually added. This is commonly referred to as the \emph{geometric} setting. However, we focus on the problem in the \emph{combinatorial} setting, where we are only given a combinatorial embedding of the graph (the clockwise order of edges around each vertex). Even in this setting, not all edges in a triangulation are flippable. Gao \etal\ \cite{GUW-GC01} showed that in every $n$-vertex triangulation at least $n-2$ edges are always flippable and that there exist some triangulations where at most $n-2$ edges are flippable. Moreover, if the triangulation has minimum degree at least 4, then they showed that there are at least $2n+3$ flippable edges and the bound is tight in certain cases. Note that by flipping an edge $e$, we transform one triangulation into another. But this triangulation could be isomorphic to the triangulation prior to the flip, or we could end up in a cycle of a few triangulations. This gives rise to the following question: Can any $n$-vertex triangulation be transformed into any other $n$-vertex triangulation through a finite sequence of flips? To our knowledge, Wagner \cite{wagner1936bemerkungenzum} was the first to address this question directly and he answered it in the affirmative. Although it is well known that the number of $n$-vertex triangulations is exponential in $n$, Wagner's inductive proof gives rise to a construction algorithm that can achieve this transformation using at most $2n^2$ edge flips. The surprising element of Wagner's proof is that he circumvents the issue of graph isomorphism by showing how to convert any given triangulation into a {\em canonical} triangulation that can be easily recognized. The canonical triangulation is the unique triangulation that consists of two dominant vertices (vertices that are adjacent to all other vertices). Unfortunately, the curse of this approach is that one may use many more flips than necessary to convert one triangulation into another. In fact, it is possible that two triangulations are one edge flip away from each other but Wagner's approach uses a quadratic number of flips to convert one into the other.

The notion of two triangulations being ``close" to each other in terms of number of flips can be expressed through a {\em flip graph}. The {\em flip graph} has as vertex set all possible $n$-vertex triangulations and two vertices in the flip graph are joined by an edge provided that the respective triangulations differ by exactly one flip. Questions about the flip operation can be viewed as questions on the flip graph. Asking whether any $n$-vertex triangulation can be converted into any other via flips is asking whether the flip graph is connected. Asking for the smallest number of flips required to convert one triangulation into another is asking for the shortest path in the flip graph between the two vertices representing the given triangulations. The maximum, minimum and average degree in the flip graph almost correspond to the maximum, minimum and average number of flippable edges. The caveat is that one needs to account for isomorphic triangulations when computing the degree of a vertex in the flip graph. One can also ask what the chromatic number of the flip graph is, whether it is hamiltonian etc. Many of these questions have been addressed in the literature~\cite{bose2009flips}. However, we focus mainly on attempts to determine the diameter of the flip graph. In other words, how many edge flips are sufficient and sometimes necessary to transform a given triangulation into any other? We go beyond merely stating the results by providing a substantial amount of detail on the proofs.

\section{Wagner's Bound}
\label{sec:wagner}

Wagner in 1936 \cite{wagner1936bemerkungenzum} first addressed the problem of determining whether one can convert a given triangulation into another via edge flips. Although his paper is entitled ``Remarks on the four-color problem'', it contains a proof that every planar graph has a straight-line embedding, defines the edge flip operation (or diagonal transformation, as Wagner calls it) and shows that any two triangulations can be transformed into each other by a finite series of edge flips before finishing with a result on the number of valid colorings of a graph. 

\begin{figure}[ht]
\begin{minipage}[t]{0.45\linewidth}
\centering
\includegraphics{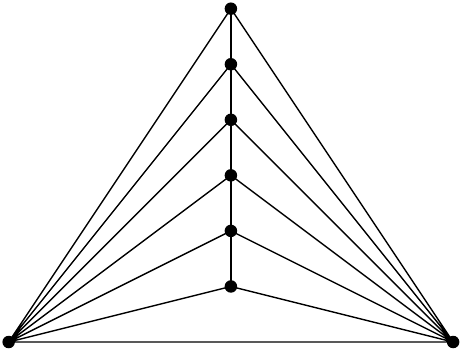}
\caption{The canonical triangulation on 8 vertices.}
\label{fig:wagner_canonical}
\end{minipage}
\hspace{0.5cm}
\begin{minipage}[t]{0.45\linewidth}
\centering
\includegraphics{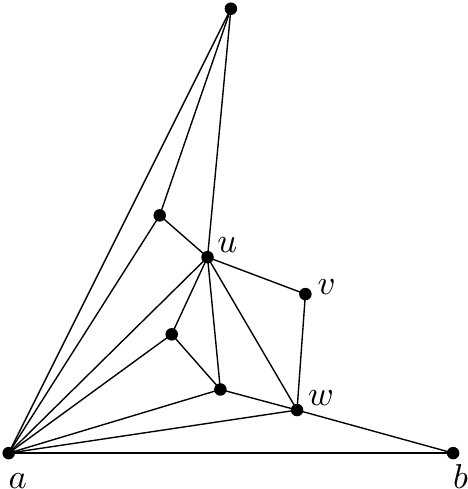}
\caption{A face $uwv$ such that $u$ and $w$ are neighbours of $a$, while $v$ is not. Flipping the edge $uw$ brings us closer to the canonical triangulation.}
\label{fig:wagner_proof}
\end{minipage}
\end{figure}

To prove that any pair of triangulations can be transformed into each other via flips, Wagner first introduces the \emph{canonical triangulation}, which is the unique triangulation with two dominant vertices (see Figure~\ref{fig:wagner_canonical}). We will denote the canonical triangulation on $n$ vertices by $\triangle_n$.

\begin{lemma}
 \label{lem:wagner}
 \from{Wagner \cite{wagner1936bemerkungenzum}, Theorem 4}
 Any triangulation on $n$ vertices can be transformed into $\triangle_n$ by a sequence of at most $n^2 - 7n + 12$ flips.
\end{lemma}
\begin{myproof}
 To transform a given triangulation into the canonical one, we fix an outer face and pick two of its vertices, say $a$ and $b$, to become the dominant vertices in the canonical triangulation. If $a$ is not adjacent to all other vertices, there exists a face $uwv$ such that $u$ and $w$ are neighbours of $a$, while $v$ is not. This situation is illustrated in Figure~\ref{fig:wagner_proof}. We flip the edge $uw$.

 In his original proof, Wagner argues that this gives a finite sequence of flips that increases the degree of $a$ by one. He simply states that this sequence is finite and does not argue why $uw$ is flippable in the first place. We provide these additional arguments below.

 We consider two cases:
 \begin{itemize}
  \item $auw$ is a face. In this case, the flip will result in the edge $av$, increasing the degree of $a$ by one. This flip is valid, as $v$ was not adjacent to $a$ before the flip.

  \item $auw$ is not a face. In this case the flip is also valid, since $auw$ forms a triangle that separates $v$ from the vertices inside. The flip does not increase the degree of $a$, but it does increase the degree of $v$ and since the number of vertices is finite, the degree of $v$ cannot increase indefinitely. Therefore, we must eventually arrive to the first case, where we increase the degree of $a$ by one.
 \end{itemize}

 Since the same strategy can be used to increase the degree of $b$ as long as it is not dominant, this gives us a sequence of flips that transforms any triangulation into the canonical one. Every vertex of a triangulation has degree at least 3, so the degree of $a$ and $b$ needs to increase by at most $n - 4$. Since we might need to increase the degree of $v$ from 2 until it is adjacent to all but one of the neighbours of $a$ or $b$, the total flip sequence has length at most
 \[ 2 \sum_{i=3}^{n-2} (i - 2) = n^2 - 7n + 12\].
\end{myproof}

By using the canonical triangulation as an intermediate form, the main result follows.

\begin{theorem}
 \from{Wagner \cite{wagner1936bemerkungenzum}, Theorem 4}
 Any pair of triangulations $T_1$ and $T_2$ on $n$ vertices can be transformed into each other by a sequence of at most $2n^2 - 14n + 24$ flips.
\end{theorem}
\begin{myproof}
 By Lemma~\ref{lem:wagner}, we have two sequences of flips, $S_1$ and $S_2$, that transform $T_1$ and $T_2$ into the canonical triangulation, respectively. Since a flip can be reversed, we can use $S_1$, followed by the reverse of $S_2$ to transform $T_1$ into $T_2$. Since both $S_1$ and $S_2$ have length at most $n^2 - 7n + 12$, the total sequence uses at most $2n^2 - 14n + 24$ flips.
\end{myproof}

A simpler and more precise proof that also gives a quadratic upper bound was given by Negami and Nakamoto \cite{negami1993diagonal}.

\begin{figure}
 \centering
 \includegraphics{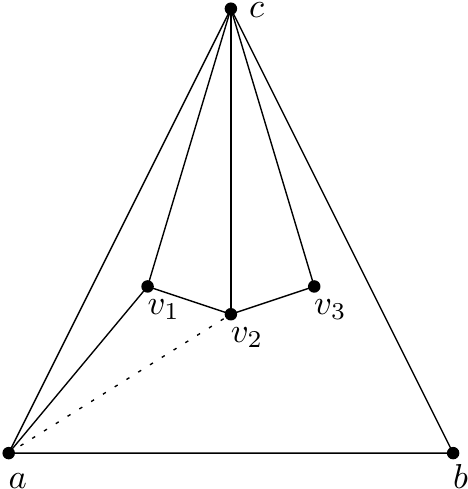}
 \caption{The exterior triangle $abc$ with the first three neighbours of $c$ in counter-clockwise order. Depending on the presence of edge $av_2$, either $cv_1$ or $cv_2$ is flipped.}
 \label{fig:wagner_negami}
\end{figure}

\begin{lemma}
 \from{Negami and Nakamoto \cite{negami1993diagonal}, Theorem 1}
 Any triangulation on $n$ vertices can be transformed into $\triangle_n$ by a sequence of $O(n^2)$ flips.
\end{lemma}
\begin{myproof}
 Let $abc$ be the outer face. Suppose we wish to make both $a$ and $b$ dominant. Instead of showing that a sequence of flips can always increase the degree of $a$ or $b$, we will show that it is always possible to find one flip that decreases the degree of $c$. Once $c$ has degree 3, the same argument can be used to find a flip that decreases the degree of $c$'s neighbour inside the triangle until it has degree 4, and so on.

 To determine which edge to flip, let $a, v_1, v_2, \dots, b$ be the neighbours of $c$ in counter-clockwise order. This situation is illustrated in Figure~\ref{fig:wagner_negami}. If $a$ and $v_2$ are not adjacent, we can flip $cv_1$ into $av_2$, reducing the degree of $c$. If $a$ and $v_2$ are adjacent, $av_2c$ forms a cycle that separates $v_1$ and $v_3$, so we can flip $cv_2$ to reduce $c$'s degree. We continue this until $c$ has degree 3, at which point we apply the same argument to reduce the degree of $c$'s remaining neighbour inside the triangle until it has degree 4. Then we continue with the neighbour of $v_1$ inside the triangle $av_1b$, and so on, until all vertices except for $a$ and $b$ have degree 3 or 4, at which point we have obtained the canonical triangulation.
\end{myproof}

\section{Komuro's Bound}
\label{sec:komuro}

Since Wagner's result, it remained an open problem whether the diameter of the flip graph was indeed quadratic in the number of vertices. Komuro \cite{komuro1997diagonal} showed that in fact the diameter was linear by proving a linear upper and lower bound. We present the argument for the upper bound in this section and discuss the lower bound in Section \ref{sec:lower}.

\begin{figure}
 \centering
 \includegraphics{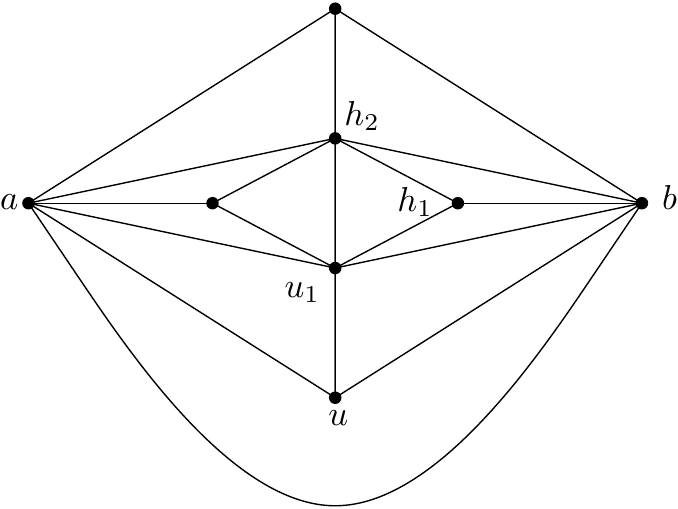}
 \caption{No single edge can be flipped to increase the degree of $a$ or $b$.}
 \label{fig:komuro_proof}
\end{figure}

Komuro used Wagner's approach of converting a given triangulation into the canonical triangulation. Given an arbitrary triangulation, the key is to bound the number of flips needed to make two vertices, say $a, b$, dominant. If there always exists one edge flip that increases the degree of $a$ or $b$ by 1, then at most $2n-8$ flips are sufficient since dominant vertices have degree $n-1$ and all vertices in a triangulation have degree at least 3. However, this is not always the case. Figure \ref{fig:komuro_proof} shows a triangulation where no single flip increases the degree of $a$ or $b$. Komuro used the following function to bound the number of flips: $d_G(a,b) = 3 \deg(a) + \deg(b)$. He showed that there always exists either one edge flip where $d_G(a,b)$ goes up by at least 1 or two edge flips where $d_G(a,b)$ goes up by at least 2. The cleverness of the function is that in some cases, two edge flips increase the degree of $a$ by 1 but decrease the degree of $b$ by 1. However, since the function increases by 2, it still increases by at least 1 per flip. Since $d_G(a,b) \leq 4n-4$, we have that $4n - 4 - d_G(a,b)$ is an upper bound on the number of flips required to make $a$ and $b$ dominant.

\begin{lemma}
 \label{lem:komuro}
 \from{Komuro \cite{komuro1997diagonal}, Lemma 2}
Let $G$ be a triangulation on $n$ vertices and let $a,b$ be any pair of adjacent vertices of $G$. Then $G$ can be transformed into the canonical triangulation $\triangle_n$ with $a$ and $b$ as dominant vertices with at most $4n-4-(3\deg(a) + \deg(b))$ edge flips. 
\end{lemma}
\begin{myproof} 
In a triangulation, every vertex must have degree at least 3. Let $uab$ be a face adjacent to $ab$. We consider two cases: $\deg(u) = 3$ and $\deg(u) > 3$. We begin with the latter. Since $\deg(u) \geq 4$, let $a, b, w_1, w_2$ be four consecutive neighbours of $u$ in counter-clockwise order. If $b$ is not adjacent to $w_2$, then flipping edge $uw_1$ increases $\deg(b)$ by 1 and thus $d_G(a,b)$ by 1. If $b$ is adjacent to $w_2$, then $ubw_2$ is a separating triangle (a cycle of length 3 whose removal disconnects the graph) that separates $a$ from $w_1$. Therefore, flipping edge $ub$ decreases $\deg(b)$ by 1 and increases $\deg(a)$ by 1. Thus, with one flip $d_G(a,b)$ increases by 2.

Now consider the case when $\deg(u) = 3$. Let $u_1$ be the unique vertex adjacent to $u, a, b$. We now have 3 cases to consider: $\deg(u_1) = 3, \deg(u_1) \geq 5, \deg(u_1) = 4$. If $\deg(u_1) = 3$, then the graph is isomorphic to $K_4$ which is $\triangle_4$. If $\deg(u_1) \geq 5$, let $a, u, b, h_1, h_2$ be five consecutive neighbours of $u_1$ in counter-clockwise order. If $b$ is not adjacent to $h_2$, then flipping edge $u_1h_1$ increases $\deg(b)$ by 1 and thus $d_G(a,b)$ by 1. If $b$ is adjacent to $h_2$, then $u_1bh_2$ is a separating triangle that separates $u, a$ from $h_1$ (see Figure \ref{fig:komuro_proof}). Therefore, flipping edge $u_1b$ and $u_1u$ decreases $\deg(b)$ by 1 and increases $\deg(a)$ by 1. Thus, with two flips $d_G(a,b)$ increases by 2.

Finally, if $\deg(u_1)=4$, then there is unique vertex $u_2$ adjacent to $a, u_1, b$. If $\deg(u_2) = 3$, the graph is isomorphic to $\triangle_5$. If $\deg(u_2) \geq 5$ we apply the same argument as when $\deg(u_1) \geq 5$. If $\deg(u_2) = 4$, we obtain another unique vertex $u_3$. This process ends with $u_{n-3}$, at which point $a$ and $b$ are dominant. 

Since $d_G(a,b)$ increases by at least 1 for one flip and at least 2 for two flips, we note that the total number of flips does not exceed $d_{\triangle_n}(a,b) - d_G(a,b) = 4n-4-(3\deg(a) + \deg(b))$ as required.

\end{myproof}

Using this lemma, Komuro proved the following theorem.

\begin{theorem}
\label{thm:komuro}
\from{Komuro \cite{komuro1997diagonal}, Theorem 1}
Any two triangulations with $n$ vertices can be transformed into each other by at most $8n-54$ edge flips if $n\geq 13$ and at most $8n-48$ edge flips if $n\geq 7$.
\end{theorem}

\begin{myproof}
Given a triangulation $G$ on $7 \leq n \leq 12$ vertices, one can prove by contradiction that either $G$ is one flip from $\triangle_n$ or there exists an edge $ab$ where both vertices have degree at least 5 implying that $d_G(a,b) \geq 20$. This gives an upper bound of $4n-24$ to convert $G$ to $\triangle_n$, which gives an upper bound of $8n-48$ to convert any triangulation to any other via the canonical triangulation. Moreover, for $n\geq 13$, either $G$ is one flip from canonical or there exists an edge $ab$ where $a$ has degree at least 6 and $b$ has degree at least 5. This means that $d_G(a,b) \geq 23$. The result follows.
\end{myproof}

\section{Mori \etal's Bound}
\label{sec:mori}

In 2001, Mori, Nakamoto and Ota \cite{mori2003diagonal} improved the bound by Komuro to $6n - 30$. They used a two-step approach by finding a short path to a strongly connected kernel, which consists of all Hamiltonian triangulations. An $n$-vertex triangulation is Hamiltonian if it contains a Hamiltonian cycle, i.e. a cycle of length $n$. The general idea of the proof is to find a fast way to make any triangulation Hamiltonian and then use the Hamiltonian cycle to decompose the graph into two outerplanar graphs. These have the following nice property.

\begin{lemma}
 \label{lem:mori_outerplanar}
 \from{Mori~\etal~\cite{mori2003diagonal}, Lemma 8 and Proposition 9}
 Any vertex $v$ in a maximal outerplanar graph on $n$ vertices can be made dominant by $n - 1 - deg(v)$ flips.
\end{lemma}
\begin{myproof}
 If $v$ is not dominant, there is a triangle $vxy$ where $xy$ is not an edge of the outer face. Then we can flip $xy$ into $vz$, where $z$ is the other vertex of the quadrilateral formed by the two triangles that share $xy$. This flip must be legal, since if $vz$ was already an edge, the graph would have $K_4$ as a subgraph, which is impossible for outerplanar graphs. Since each such flip increases the degree of $v$ by one, $n - 1 - deg(v)$ flips are both necessary and sufficient.
\end{myproof}

With this property, Mori~\etal~showed that it is possible to quickly transform any Hamiltonian triangulation into the canonical form by decomposing it along the Hamiltonian cycle into two outerplanar graphs. Interestingly, this exact approach was already used 10 years earlier by Sleator, Tarjan and Thurston~\cite{sleatortq1992short} to prove a $\Theta(n \log n)$ bound on the diameter of the flip graph when the vertices are labelled. Note that this was even before the linear bound by Komuro that was discussed in the previous section. However, they did not state their result in terms of unlabelled triangulations and it seems that both Komuro and Mori~\etal~were unaware of this earlier work.

\begin{figure}[ht]
 \centering
 \includegraphics{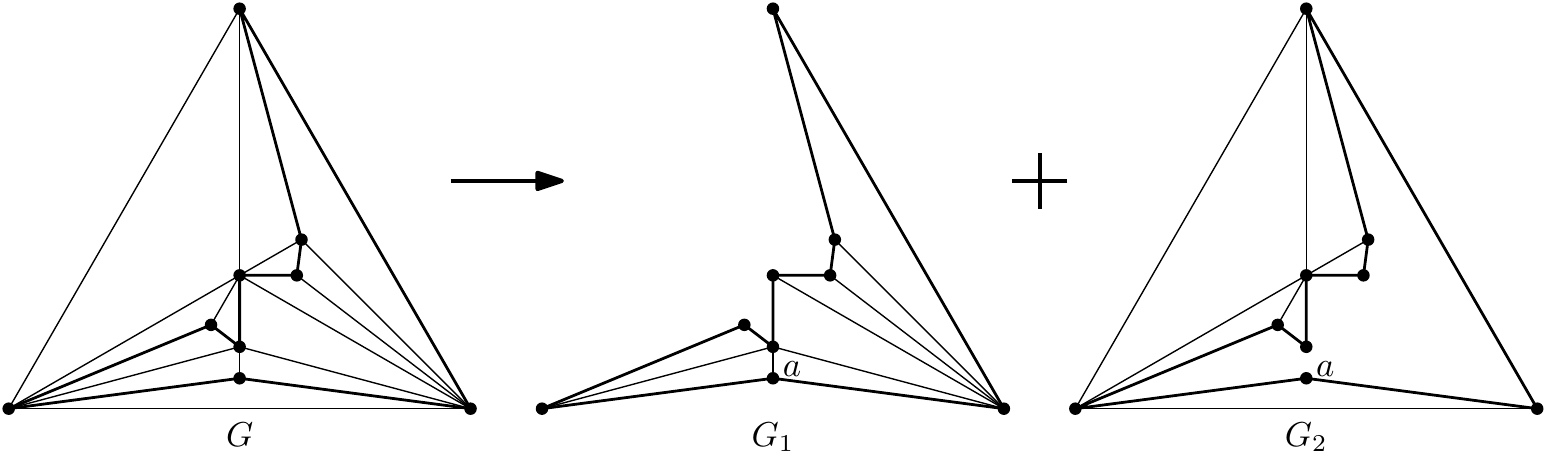}
 \caption{The decomposition of a Hamiltonian graph $G$ into two outerplanar graphs $G_1$ and $G_2$. The vertex $a$ has degree 2 in $G_2$.}
 \label{fig:mori_decomposition}
\end{figure}

\begin{theorem}
 \label{thm:mori_hamiltonian}
 \from{Mori~\etal~\cite{mori2003diagonal}, Proposition 9}
 Any Hamiltonian triangulation on $n$ vertices can be transformed into $\triangle_n$ by at most $2n - 10$ flips, preserving the existence of Hamiltonian cycles.
\end{theorem}
\begin{myproof}
 Given a Hamiltonian triangulation $G$ with Hamiltonian cycle $C$, we can decompose it into two outerplanar graphs $G_1$ and $G_2$, such that each contains $C$ and all edges on one side of $C$. This is illustrated in Figure~\ref{fig:mori_decomposition}. Let $a$ be a vertex of degree 2 in $G_2$. We are going to make $a$ dominant in $G_1$. Since $G$ is 3-connected and $a$ has no additional neighbours in $G_2$, the degree of $a$ in $G_1$ is at least 3. Thus by Lemma~\ref{lem:mori_outerplanar}, we can make $a$ dominant by at most $n - 4$ flips. Each of these flips is valid, as $a$ is not connected to anyone in $G_2$, except for its neighbours on $C$.

 Now consider the subgraph $G_2' = G_2 \setminus \{a\}$. Since $a$ has degree 2 in $G_2$, $G_2'$ is still outerplanar, so by applying Lemma~\ref{lem:mori_outerplanar} again we can make a vertex of $G_2'$ dominant as well, which gives us the canonical triangulation. Since $G_2'$ has $n - 1$ vertices and it always has a vertex of degree at least 4 (provided that $n \geq 6$), we need at most $n - 6$ flips for this. Since we did not flip any of the edges on $C$, the theorem follows.
\end{myproof}

This shows that the Hamiltonian triangulations are closely connected, so all we need to figure out is how we can quickly make a triangulation Hamiltonian. Here, we turn to an old result by Whitney~\cite{whitney1931theorem} that shows that all 4-connected triangulations are Hamiltonian. Since a triangulation is 4-connected if and only if it does not have any separating triangles (cycles of length 3 whose removal disconnects the graph), by removing all separating triangles from a triangulation, we make it 4-connected and therefore Hamiltonian. Fortunately, separating triangles are easy to remove using flips, as the following lemmas show.

\begin{lemma}
 \label{lem:mori_flip}
 \from{Mori~\etal~\cite{mori2003diagonal}, Lemma 11}
 In a triangulation with $n \geq 6$ vertices, flipping any edge of a separating triangle $D = abc$ will remove that separating triangle. This never introduces a new separating triangle, provided that the selected edge belongs to multiple separating triangles or none of the edges of $D$ belong to multiple separating triangles.
\end{lemma}
\begin{myproof}
 Since $D$ is separating and the newly created edge connects a vertex on the inside to a vertex on the outside, the flip is always legal. Since the flip removes an edge of $D$, it is no longer a separating triangle. Now suppose that we flipped $ab$ to a new edge $xy$ and introduced a new separating triangle $D'$. Then $D'$ must be $xyc$.
 But since $n \geq 6$ and our construction so far uses only 5 vertices, one of the faces $ayc, byc, axc, \text{ or } bcx$ must be a separating triangle as well. This means that either $ac$ or $bc$ is an edge that belongs to multiple separating triangles, while $ab$ only belongs to $D$, which contradicts the choice of $ab$.
\end{myproof}

\begin{lemma}
 \label{lem:mori_4connected}
 \from{Mori~\etal~\cite{mori2003diagonal}, Lemma 11}
 Any triangulation on $n$ vertices can be made 4-connected by at most $n - 4$ flips.
\end{lemma}
\begin{myproof}
 We will show that a triangulation can have at most $n - 4$ separating triangles, the result follows by Lemma~\ref{lem:mori_flip}. The proof is by induction on $n$. For the base case, let $n = 4$. Then our graph must be $K_4$, which has no separating triangles as required. For the induction we can assume that our graph $G$ has a separating triangle $T$ which partitions $G$ into two components $G_1$ and $G_2$. By induction, $G_1$ and $G_2$ have at most $n_1 - 4$ and $n_2 - 4$ separating triangles, where $n_1$ and $n_2$ are the number of vertices in $G_1$ and $G_2$, respectively, including the vertices of $T$. Therefore $G$ can have at most $n_1 - 4 + n_2 - 4 + 1 = (n_1 + n_2 - 3) - 4 = n - 4$ separating triangles.
\end{myproof}

Now we can prove the main result.

\begin{theorem}
 \from{Mori~\etal~\cite{mori2003diagonal}, Theorem 4}
 Any two triangulations on $n$ vertices can be transformed into each other by at most $6n - 30$ flips.
\end{theorem}
\begin{myproof}
 The connection between Lemma~\ref{lem:mori_4connected} and Theorem~\ref{thm:mori_hamiltonian} is an old proof by Whitney~\cite{whitney1931theorem} that any 4-connected triangulation is Hamiltonian. Therefore we can transform any triangulation into the canonical form by at most $n - 4 + 2n - 10 = 3n - 14$ flips. By looking carefully at the proof of Theorem~\ref{thm:mori_hamiltonian}, we see that if the graph is 4-connected, the first vertex (vertex $a$) is guaranteed to have degree at least 4, which brings the bound down to $3n - 15$ flips to the canonical triangulation and $6n - 30$ flips between any pair of triangulations.
\end{myproof}

\section{Bose \etal's Bound}
\label{sec:bose}

Mori's bound on the number of flips required to remove all separating triangles in a triangulation is not tight. Recently, Bose~\etal~\cite{bose2011making} showed that $(3n - 6)/5$ flips suffice and are sometimes necessary to make a triangulation 4-connected. We give a summary of the proof below, the full details can be found in the original paper. The matching lower bound is described in Section~\ref{sec:lower}. First, we divide the edges of the graph into two types; edges that belong to separating triangles and edges that do not. The latter are called \emph{free edges} and they have the following nice property.

\begin{lemma}
 \label{lem:freeedge}
 \from{Bose~\etal~\cite{bose2011making}, Lemma 2}
 In a triangulation $T$, every vertex $v$ of a separating triangle $D$ is incident to at least one free edge inside $D$.
\end{lemma}
\begin{myproof}
 Consider one of the edges of $D$ incident to $v$. Since $D$ is separating, its interior cannot be empty and since $D$ is part of $T$, there is a triangular face inside $D$ that uses this edge. Now consider the other edge $e$ of this face that is incident to $v$.

 The remainder of the proof is by induction on the number of separating triangles contained in $D$. For the base case, assume that $D$ does not contain any other separating triangles. Then $e$ must be a free edge and we are done.

 For the induction step, there are two further cases. If $e$ does not belong to a separating triangle, we are again done, so assume that $e$ belongs to a separating triangle $D'$. Since $D'$ is itself a separating triangle contained in $D$ and containment is transitive, the number of separating triangles contained by $D'$ must be strictly smaller than that of $D$. Since $v$ is also a vertex of $D'$, our induction hypothesis tells us that there is a free edge incident to $v$ inside $D'$. Since $D'$ is contained in $D$, this edge is also inside $D$.
\end{myproof}

Separating triangles can be contained in other separating triangles. A \emph{deepest} separating triangle is one that is contained in the maximum number of separating triangles. We will remove all separating triangles by repeatedly flipping an edge of a deepest separating triangle.

\begin{theorem}
 \label{thm:bose_4-connected}
 \from{Bose~\etal~\cite{bose2011making}, Theorem 3}
 A triangulation on $n \geq 6$ vertices can be made 4-connected using at most $\lfloor(3n - 6)/5\rfloor$ flips.
\end{theorem}
\begin{myproof}
 We prove this using a simple charging scheme; we place one coin on every edge at the start and charge 5 coins per flip. This guarantees that we perform at most $\lfloor(3n - 6)/5\rfloor$ flips. To prove that we can actually charge 5 coins per flip, we need two invariants:
 \begin{shortitemize}
  \item Every edge of a separating triangle has a coin.
  \item Every vertex of a separating triangle has an incident free edge that is inside the triangle and that has a coin.
 \end{shortitemize}
 We restrict ourselves to flipping edges of a deepest separating triangle $D$. This gives us four types of edges we can charge:

  \smallskip
  \noindent \textbf{Type 1 (\includegraphics{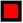})}. The flipped edge $e$. The flip removes all separating triangles that $e$ belongs to and does not introduce any new ones, so both invariants are still satisfied if we remove $e$'s coin.

  \smallskip
  \noindent \textbf{Type 2 (\includegraphics{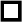})}. An edge $e$ of $D$ that is not shared with any other separating triangle. Again, the flip removes the separating triangle that $e$ belongs to and does not introduce any new ones, so we can safely charge $e$'s coin.

  \smallskip
  \noindent \textbf{Type 3 (\includegraphics{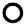})}. A free edge $e$ of a vertex of $D$ that is not shared with any containing separating triangle. Since the flip removed $D$ and $e$ is not incident to a vertex of another separating triangle that contains it, it is no longer required to have a coin to satisfy the second invariant.

  \smallskip
  \noindent \textbf{Type 4 (\includegraphics{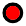})}. A free edge $e$ incident to a vertex $v$ of $D$, where $v$ is an endpoint of an edge $e'$ of $D$ that is shared with a non-containing separating triangle $B$, provided that we flip $e'$. Any separating triangle that contains $D$ but not $B$ must share $e'$ and is therefore removed by the flip. So every separating triangle after the flip that shares $v$ and contains $D$ also contains $B$. Since the second invariant requires only one free edge with a coin for each vertex, we can safely charge the free edge inside $D$, as long as we do not charge the one in $B$.
 
 \medskip
 To decide which edge we flip and how we pay for each flip, we distinguish five cases for $D$, based on the number of edges shared with other separating triangles and whether any of these triangles contain $D$. These cases are illustrated in Figures~\ref{fig:bose_caseA}, \ref{fig:bose_caseC}, and \ref{fig:bose_caseB}.

 \begin{figure}[ht]
  \centering
  \includegraphics{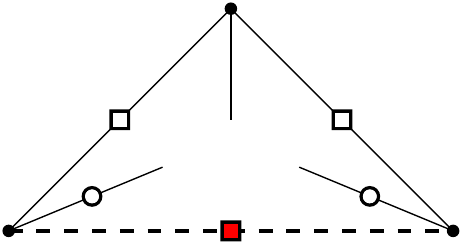}
  \caption{The edges that are charged if the deepest separating triangle does not share any edges with other separating triangles. The flipped edge is dashed and the charged edges are marked with shaded boxes (Type~1), white boxes (Type~2), white disks (Type~3) or shaded disks (Type~4).}
  \label{fig:bose_caseA}
 \end{figure}

 \smallskip
  \noindent Case 1. $D$ does not share any edges with other separating triangles (Figure~\ref{fig:bose_caseA}). In this case, we flip any one of $D$'s edges and charge all of them. Since $D$ can share at most one vertex with a containing triangle, we charge the remaining two coins from two free edges, each incident to one of the other two vertices.

 \begin{figure}[ht]
  \centering
  \includegraphics{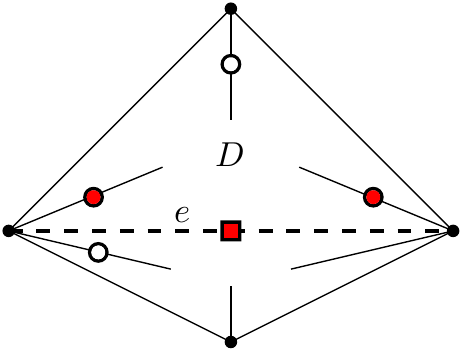}
  \caption{The edges that are charged if the deepest separating triangle only shares edges with non-containing separating triangles.}
  \label{fig:bose_caseC}
 \end{figure}

 \smallskip
  \noindent Case 2. $D$ does not share any edge with a containing separating triangle, but shares one or more edges with non-containing separating triangles (Figure~\ref{fig:bose_caseC}). In this case, we flip one of the shared edges $e$. We charge $e$ and two free edges inside $D$ that are incident to the vertices of $e$. Now consider the quadrilateral formed by $D$ and the non-containing separating triangle that shares $e$ with $D$. Since $D$ does not share an edge with a containing separating triangle, at most two vertices of this quadrilateral can be shared with containing separating triangles. Therefore we charge two free edges, each incident to one of the other two vertices, for the last two coins.
  
 \begin{figure*}[ht]
  \centering
  \includegraphics{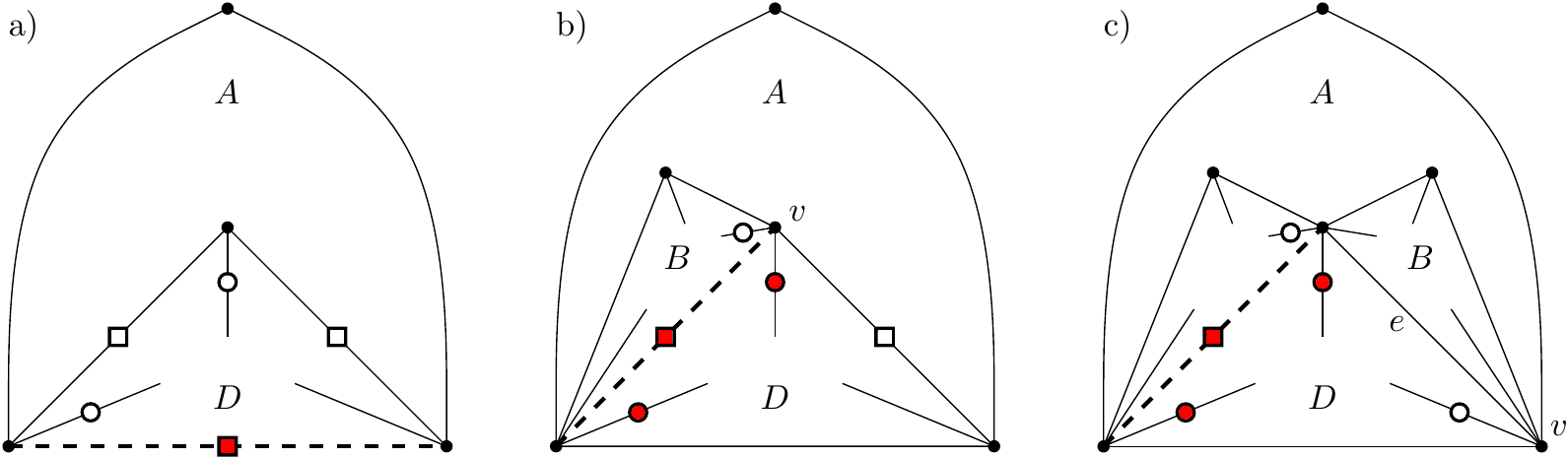}
  \caption{The edges that are charged if the deepest separating triangle shares an edge with a containing triangle.}
  \label{fig:bose_caseB}
 \end{figure*}

 \smallskip
  \noindent Case 3. $D$ shares an edge with a containing triangle $A$ and does not share the other edges with any separating triangle (Figure~\ref{fig:bose_caseB}a). In this case, we flip the shared edge and charge all of $D$'s edges. The vertex of $D$ that is not shared with $A$ cannot be shared with any containing triangle and at most one of the vertices of the shared edge can be shared with containing separating triangles that are not removed by the flip, so we charge two free edges incident to the unshared vertices for the remaining coins.

 \smallskip
  \noindent Case 4. $D$ shares an edge with a containing triangle $A$ and one other edge with a non-containing separating triangle $B$ (Figure~\ref{fig:bose_caseB}b). In this case, we flip the edge that is shared with $B$. Let $v$ be the vertex of $D$ that is not shared with $A$. We charge the flipped edge, the unshared edge of $D$ and two free edges inside $D$ that are incident to the vertices of the flipped edge. We charge the last coin from a free edge in $B$ that is incident to $v$. Every separating triangle that contains $D$ must contain $B$ as well and since $D$ is deepest, $B$ must be deepest too. Therefore $v$ cannot be shared with any other separating triangle that contains this free edge and we can safely charge it.

 \smallskip
  \noindent Case 5. $D$ shares one edge with a containing triangle $A$ and the other two with non-containing separating triangles (Figure~\ref{fig:bose_caseB}c). In this case we flip one of the edges shared with non-containing triangles. The charged edges are identical to the previous case, except that there is no unshared edge any more. Instead, we charge the last free edge in $D$. This is allowed, as there is still an uncharged free edge incident to this vertex inside the separating triangle $B$ whose edge we do not flip.
 
 This shows that we can charge 5 coins for every flip, while maintaining the invariants. As long as our triangulation has a separating triangle, we can always find a deepest separating triangle $D$, which must fit into one of the cases above. This gives us an edge of $D$ to flip and five edges to charge, each of which is guaranteed by the invariants to have a coin. Therefore the process stops only after all separating triangles have been removed.
\end{myproof}

By combining this result with the bound for 4-connected triangulations by Mori~\etal, we obtain a better bound on the diameter of the flip graph.

\begin{coro}
 Any pair of triangulations on $n$ vertices can be transformed into each other by at most $5.2 n - 24.4$ flips.
\end{coro}
\begin{myproof}
 Mori~\etal~\cite{mori2003diagonal} showed that any two 4-connected triangulations can be transformed into each other by at most $4n - 22$ flips. By Theorem~\ref{thm:bose_4-connected}, we can make a triangulation 4-connected using at most $\lfloor (3n - 6)/5 \rfloor$ flips. Hence, we can transform any triangulation into any other using at most $2 \cdot (3n - 6)/5 + 4n - 22 = 5.2 n - 24.4$ flips.
\end{myproof}

\section{Lower Bounds}
\label{sec:lower}

The best known lower bound on the diameter of the flip graph is by Komuro~\cite{komuro1997diagonal} and is based on the maximum degree of the vertices in the graph.

\begin{theorem}
 \from{Komuro~\cite{komuro1997diagonal}, Theorem 5} Let $G$ be a triangulation on $n$ vertices. Then at least $2n - 2 \Delta(G) - 3$ flips are needed to transform $G$ into the canonical triangulation, where $\Delta(G)$ denotes the maximum degree of $G$.
\end{theorem}
\begin{myproof}
 Let $a$ and $b$ be the two vertices of degree $n - 1$ in the canonical triangulation. Each flip increases the degree in $G$ of either $a$ or $b$ by at most one. The only possible exception is the flip that creates the edge $ab$, which increases the degree of both vertices by one. Since the initial degree of $a$ and $b$ is at most $\Delta(G)$, we need at least $2(n - 1 - \Delta(G)) - 1 = 2n - 2\Delta(G) - 3$ flips.
\end{myproof}

Since there are triangulations that have maximum degree 6, this gives a lower bound of $2n - 15$ flips. It is interesting that one of the triangulations in the lower bound is the canonical form. This implies that either the lower bound is very far off, or the canonical triangulation is a bad choice of intermediate triangulation. It also means that as long as we use this canonical form, the best we can hope for is an upper bound of $4n - 30$ flips. Komuro also gave a lower bound on the number of flips required to transform between any pair of triangulations, again based on the degrees of the vertices.

\begin{theorem}
 \from{Komuro~\cite{komuro1997diagonal}, Theorem 4} Let $G$ and $G'$ be triangulations on $n$ vertices. Let $v_1, \dots, v_n$ and $v_1', \dots, v_n'$ be the vertices of $G$ and $G'$, respectively, ordered by increasing degree. Then at least $\frac{1}{4}D(G, G')$ flips are needed to transform $G$ into $G'$, where $D(G, G') = \sum_{i=1}^n |deg(v_i) - deg(v_i')|$.
\end{theorem}
\begin{myproof}
 Let $\sigma$ be a mapping between the vertices of $G$ and $G'$ and suppose we transform $G$ into $G'$ using flips, such that $v_i \in G$ becomes $v_{\sigma(i)}' \in G'$. Since every flip changes the degree of a vertex by one, we need at least $|deg(v_i) - deg(v_{\sigma(i)}')|$ flips to obtain the correct degree for $v_{\sigma(i)}'$. However, each flip affects the degrees of 4 vertices, giving a bound of $\frac{1}{4}\sum_{i=1}^n |deg(v_i) - deg(v_{\sigma(i)}')|$ flips. Our actual lower bound is the minimum of this bound over all mappings $\sigma$. Mapping every vertex to a vertex with the same rank when ordered by degree (i.e. $\sigma(i) = i$) achieves this minimum.
\end{myproof}

There are stricter lower bounds if we restrict ourselves to the approach used by Mori~\etal. First, since there are Hamiltonian triangulations with maximum degree 6, the $2n - 15$ lower bound holds for the transformation of a Hamiltonian triangulation to the canonical one as well. The technique used by Mori~\etal\ uses at most $2n - 10$ flips, so it is only 5 flips removed from the lower bound. There is also a tight lower bound by Bose~\etal\ for the number of flips required to make a triangulation 4-connected. 

\begin{figure}[ht]
 \centering
 \includegraphics{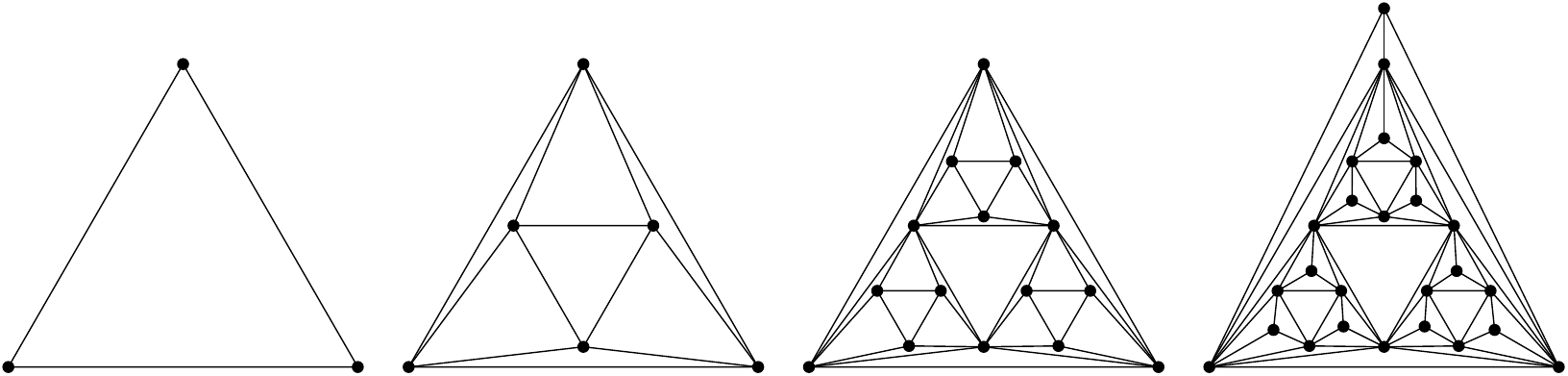}
 \caption{Recursive construction of a triangulation with a large number of edge-disjoint separating triangles.}
 \label{fig:lower_sierpinsky}
\end{figure}

\begin{theorem}
 \from{Bose~\etal\ \cite{bose2011making}, Theorem 5} There are triangulations that require $\lceil(3n - 10)/5\rceil$ flips to make them 4-connected.
\end{theorem}
\begin{myproof}
 This bound is based on the recursive construction illustrated in Figure~\ref{fig:lower_sierpinsky}. It starts with a single triangle, adds an inverted triangle and connects each vertex to both vertices of the opposing edge. Then it recurses on three of the new triangles incident to the original triangle vertices. In the final step of the recursion, instead of an inverted triangle, a single vertex is added and connected to all three triangle vertices. One vertex is also added to the exterior face, so the original triangle becomes separating as well. This generates a triangulation with $(3n - 10)/5$ edge-disjoint separating triangles. Since removing all separating triangles is the only way to make a graph 4-connected, this requires at least $\lceil(3n - 10)/5\rceil$ flips. This differs less than a single flip from the upper bound of $\lfloor(3n - 6)/5\rfloor$ and since the number of flips is necessarily integer, these bounds are tight.
\end{myproof}

This gives a strong indication that we have reached the limit of Mori~\etal's approach of converting a triangulation to the canonical form by first making it 4-connected. However, 4-connectedness is only a sufficient condition for Hamiltonicity, not a necessary one. There are a lot of examples (especially small ones) of triangulations that are Hamiltonian, while not being 4-connected. Therefore it might be possible to make a triangulation Hamiltonian with fewer flips. The best known lower bound on this number is by Aichholzer, Huemer and Krasser~\cite{oswin08}.

\begin{theorem}
 \from{Aichholzer~\etal\ \cite{oswin08}, Corollary 6} There are triangulations on $n$ vertices that require at least $(n - 8)/3$ flips to make them Hamiltonian.
\end{theorem}
\begin{myproof}
 Take any triangulation on $k \geq 5$ vertices, then add a new vertex to each face and connect it to all three vertices of the face. The new triangulation has $n = 3k - 4$ vertices. Call the original vertices white and the new vertices black. Since every black vertex is adjacent only to white vertices, a Hamilton cycle would have to pass through at least one white vertex between every pair of black vertex. However, there are $2k - 4$ black vertices and only $k$ white vertices, so no such cycle can exist. Furthermore, a single flip can reduce the number of black components by at most one. Therefore at least $k - 4 = (n - 8)/3$ flips are required to make this triangulation Hamiltonian.
\end{myproof}

This shows that there is still a small improvement possible, but we most likely need new techniques to reduce the gap between the upper and lower bounds further.

\section{Conclusions}
\label{sec:conclusion}

We presented a comprehensive overview including full proofs of the following problem: Given two $n$-vertex triangulations, how many edge flips are necessary and sufficient to transform one triangulation into the other? Currently, the best known upper bound is $5.2 n - 24.4$ edge flips, while for the lower bound, there exist pairs of triangulations that require at least $2n - 15$ edge flips. There remain a number of open problems in this area; we outline a few of them here. The obvious one is to reduce the gap between the upper and lower bound. There are strong indications, as outlined in Section~\ref{sec:lower}, that the upper bound cannot be improved much further with the current techniques. Moreover, it seems counter-intuitive that the canonical triangulation used for all the upper bounds is actually one of the triangulations used to prove the best known lower bound. We believe that a new approach will be needed to improve the current upper bound. 

The main graph theoretic property used to prove the lower bounds is vertex degree. The best known lower bound is achieved by showing that converting a triangulation whose maximum degree is 6 into one with two dominant vertices requires at least $2n-15$ flips. We feel that other graph theoretic structures will have to be exploited to improve the lower bound, although the question of graph isomorphism is always lurking beneath the surface.

The main open problem in this area is to try to find a way to determine what is the smallest number of flips needed to convert a given triangulation into another. Currently, all of the known approaches can be made to use at least a linear number of flips even when two triangulations differ by one flip. The problem is that all of the approaches first convert a given triangulation into a canonical one. Any algorithm that uses a number of flips that is sensitive to the smallest number of required flips would be a major step forward and could lead to a better understanding of this problem.

\bibliography{papers}{}

\begin{thebibliography}{10}

\bibitem{oswin08}
{\sc Aichholzer, O., Huemer, C., and Krasser, H.}
\newblock Triangulations without pointed spanning trees.
\newblock {\em Comput. Geom. 40}, 1 (2008), 79--83.

\bibitem{bose2009flips}
{\sc Bose, P., and Hurtado, F.}
\newblock Flips in planar graphs.
\newblock {\em Comput. Geom. 42}, 1 (2009), 60--80.

\bibitem{bose2011making}
{\sc Bose, P., Jansens, D., van Renssen, A., Saumell, M., and Verdonschot, S.}
\newblock Making triangulations 4-connected using flips.
\newblock In {\em Proceedings of the 23rd Canadian Conference on Computational
  Geometry\/} (2011), pp.~241--247.
\newblock A full version of this paper can be found at arXiv:1110.6473.

\bibitem{GUW-GC01}
{\sc Gao, Z., Urrutia, J., and Wang, J.}
\newblock Diagonal flips in labelled planar triangulations.
\newblock {\em Graphs Combin. 17}, 4 (2001), 647--657.

\bibitem{komuro1997diagonal}
{\sc Komuro, H.}
\newblock The diagonal flips of triangulations on the sphere.
\newblock {\em Yokohama Math. J. 44}, 2 (1997), 115--122.

\bibitem{mori2003diagonal}
{\sc Mori, R., Nakamoto, A., and Ota, K.}
\newblock Diagonal flips in {H}amiltonian triangulations on the sphere.
\newblock {\em Graphs Combin. 19}, 3 (2003), 413--418.

\bibitem{negami1993diagonal}
{\sc Negami, S., and Nakamoto, A.}
\newblock Diagonal transformations of graphs on closed surfaces.
\newblock {\em Sci. Rep. Yokohama Nat. Univ. Sect. I Math. Phys. Chem.}, 40
  (1993), 71--97.

\bibitem{sleatortq1992short}
{\sc Sleator, D.~D., Tarjan, R.~E., and Thurston, W.~P.}
\newblock Short encodings of evolving structures.
\newblock {\em SIAM J. Discrete Math. 5}, 3 (1992), 428--450.

\bibitem{wagner1936bemerkungenzum}
{\sc Wagner, K.}
\newblock {Bemerkungen zum vierfarbenproblem}.
\newblock {\em Jahresber. Dtsch. Math.-Ver. 46\/} (1936), 26--32.

\bibitem{whitney1931theorem}
{\sc Whitney, H.}
\newblock A theorem on graphs.
\newblock {\em Ann. of Math. (2) 32}, 2 (1931), 378--390.

\end{thebibliography}
\bibliographystyle{acm}

\end{document}